\documentclass[11pt]{article}

\newcommand{\rom}[1]{\mathrm{#1}}

\usepackage{amsmath,amssymb,graphicx}
\usepackage{hyperref}

\usepackage{multicol,color,longtable}

\definecolor{darkred}{rgb}{0.65,0.15,0}
\hypersetup{pdfborder={0 0 0},colorlinks=true,urlcolor=darkred,citecolor=blue,linkcolor=darkred,linktocpage=true}

\definecolor{AV}{rgb}{0.65,0.0,0}
\definecolor{AK}{rgb}{0,0,1}
\definecolor{DK}{rgb}{0.6,0.4,0}

\newcommand{\nn}{\nonumber}

\newcommand{\reals}{\mathbb{R}}

\newcommand{\cV}{\mathcal{V}}

\newcommand{\cM}{\mathcal{M}}

\newcommand{\id}{1\!\!1}

\newcommand \RR{{\mathbb{R}}}

\newcommand\be{\begin{equation}}
\newcommand\ee{\end{equation}}

\newcommand\bea{\begin{eqnarray}}
\newcommand\eea{\end{eqnarray}}

\newcommand{\beq}{\begin{eqnarray}}
\newcommand{\eeq}{\end{eqnarray}}

\def \RR{{\mathbb{R}}}

\newcommand{\h}{\eta}

\makeatletter

\@addtoreset{equation}{section}
\makeatother

\begin{document}

\thispagestyle{empty}

{\flushright {AEI-2014-047}\\[15mm]}

\begin{center}
{\LARGE \bf An Inverse Scattering Construction of \\[3mm] the JMaRT Fuzzball}\\[10mm]

\vspace{8mm}
\normalsize
{\large  Despoina Katsimpouri${}^{1}$, Axel Kleinschmidt${}^{1,2}$\\[2mm] and Amitabh Virmani${}^{1,3,4}$}

\vspace{10mm}
${}^1${\it Max-Planck-Institut f\"{u}r Gravitationsphysik (Albert-Einstein-Institut)\\
Am M\"{u}hlenberg 1, DE-14476 Potsdam, Germany}
\vskip 1 em
${}^2${\it International Solvay Institutes\\
ULB-Campus Plaine CP231, BE-1050 Brussels, Belgium}
\vskip 1 em
${}^3${\it Institute of Physics\\
Sachivalaya Marg, Bhubaneshwar, Odisha, India 751005}
\vskip 1 em
${}^4${\it
Kavli Institute for Theoretical Physics China \\ Zhong Guan Cun East Street 55, Beijing, China  100190}
\vspace{15mm}

\hrule

\vspace{10mm}

\begin{tabular}{p{12cm}}
{\small
We present an inverse scattering construction in STU supergravity of the two-charge single-rotation JMaRT fuzzball. The key element in our construction is the fact that with appropriate changes in the parameters, the JMaRT fuzzball can be smoothly connected to the Myers--Perry instanton.
}
\end{tabular}
\vspace{10mm}
\hrule
\end{center}

\newpage
\setcounter{page}{1}

\tableofcontents

\section{Introduction}

Finding exact solutions of gravity or supergravity is of central importance for a proper understanding of the Einstein field equations. In view of Mathur's fuzzball proposal~\cite{Mathur:2005zp}, the need for 
exact smooth solutions becomes even more pronounced as large families of smooth solutions are conjectured to account for the entropy of  black holes. Only a handful of non-extremal examples of such families are known \cite{JMaRT, Bena:2009qv, Bobev:2009kn, Giusto:2007tt, Banerjee:2014hza};  the most notable of these are the so-called JMaRT~\cite{JMaRT} and running-Bolt solutions~\cite{Bena:2009qv,Bobev:2009kn}, see for example~\cite{Bena:2013dka} for a recent review. It would be desirable to have more examples to scrutinise
the fuzzball proposal more thoroughly.

Solution generating techniques in (super-)gravity have a long history, see for example the textbook~\cite{Stephani:2003tm}. Supposing $D-2$ commuting Killing vectors in $D$-dimensional gravity one can rewrite the Einstein
field equations in terms of an integrable linear system that is amenable to inverse scattering techniques. The most widely used such linear system is that of Belinski--Zakharov~\cite{BZ1,BZ2,BV} where a seed solution is
dressed by so-called soliton transformations. This method is very effective but presently only applicable to pure $D=4$ and $D=5$ Einstein gravity~\cite{BV,Pomeransky:2005sj}. An alternative linear system was given by
Breitenlohner--Maison~\cite{BM} that brings the underlying affine group symmetry to the fore. In~\cite{BMnotes} Breitenlohner and Maison (BM) showed how to use their linear system to generate black hole solutions in gravity. 
This method and the relation between the two linear systems was reviewed in~\cite{KKV} and applied in~\cite{KKV2} to STU supergravity~\cite{STU1}.

In the present paper we will show how to fit the JMaRT solution into the BM linear system. More precisely, we will recover the two-charge JMaRT solution from an appropriate inverse scattering construction. Our construction
also allows us to study the rod structure~\cite{Harmark:2004rm, Hollands:2007aj} of the JMaRT solution in detail. 

Generating solutions using the BM linear system requires an appropriate meromorphic monodromy matrix $\cM(w)$ depending on a spectral parameter $w$.  The physical parameters of the solutions are encoded in the positions of the 
poles of $\cM(w)$ and in the residues at these poles. Recovering the physical solution requires factorising the monodromy matrix in a specific way~\cite{BM,KKV} that is reviewed below. For STU supergravity, the monodromy
matrix is an element of $\mathrm{SO}(4,4)$ and the factorisation of $\cM(w)$ can be reduced to a purely algebraic problem~\cite{KKV2}.

For the JMaRT solution we first construct the monodromy matrix of the five-dimensional Myers--Perry instanton\footnote{To the best of our knowledge this instanton has not appeared in the literature before but is straightforward 
to construct using standard techniques.}. This is the Euclidean analogue of the Lorentzian over-rotating Myers--Perry black hole that served as a basis for the original JMaRT construction. The construction of the Myers--Perry instanton
is the genuine new element in the analysis. It is then charged up using standard methods to arrive at the two-charge JMaRT.

The plan of the article is as follows. In section~\ref{sec:prelim}, we first provide some background on the linear system and inverse scattering method used in order to provide a self-contained presentation of the construction. 
The detailed procedure of obtaining the JMaRT fuzzball through inverse scattering and the Myers--Perry instanton is then discussed in section~\ref{sec:sugra}. Section~\ref{sec:rods} contains the discussion of the rod structure and
section~\ref{sec:concl} some concluding remarks. In appendix~\ref{app:redux} we provide the necessary details and conventions on the theory studied in various dimensions and the relation of the Euclidean and Lorentzian theories in
$D=5$ embedded in the STU model as uplifts. In appendix~\ref{app:details} certain detailed intermediate expressions are given.

\section{Brief review of the inverse scattering formalism}
\label{sec:prelim}

It is well-known that STU supergravity reduced to three dimensions exhibits a global $\mathrm{SO}(4,4)$ symmetry~\cite{STU1}  and this symmetry was used
in the construction of many interesting charged solutions of the theory, see for instance~\cite{Cvetic:1996kv, CY5d, Chow:2013tia}. Together with the global $\mathrm{SO}(4,4)$ symmetry there
is a local symmetry given by a maximal subgroup $\mathrm{K}$ of $\mathrm{G}=\mathrm{SO}(4,4)$ fixed by an involution. The precise signature and embedding of $\mathrm{K}$ depends on the way 
the dimensional reduction to three dimensions is performed. In the case of stationary solutions that we are considering and that can be uplifted to $D=6$ as described in detail in appendix~\ref{app:redux},
one has $\mathrm{K}=\mathrm{SO}(2,2)\times\mathrm{SO}(2,2)$. 

The groups $\mathrm{G}$ and $\mathrm{K}$ here are defined as those real $(8\times 8)$-matrices that satisfy
\begin{subequations}
\label{GK}
\begin{align}
\mathrm{G} &= \mathrm{SO}(4,4) = \left\{ g\,|\, g^T\eta g =\eta \right\}\\
\mathrm{K} &= \mathrm{SO}(2,2)\times\mathrm{SO}(2,2) = \left\{ g\in \mathrm{SO}(4,4)\,|\, g^T\eta' g =\eta' \right\}
\end{align}
\end{subequations}
with the invariant metrics
\begin{align}
\label{invmetrics}
 \eta=\begin{pmatrix}0_4&\id_4\\\id_4 &0_4\end{pmatrix},\quad
\eta'=\textrm{diag}(+,-,-,+,+,-,-,+).
\end{align}
On the Lie algebra $\mathrm{Lie}(\mathrm{G})$ we have the anti-involution (called generalized transpose)
\begin{align}
\label{sharp}
X^\sharp = \eta' X^T \eta'
\end{align}
under which elements of $\mathrm{Lie}(\mathrm{K})$ are anti-symmetric. We define the same operation on arbitrary $(8\times 8)$-matrices. With this definition, elements $k\in \mathrm{K}$ satisfy $k^\sharp k=\id$. Note that the $\eta'$ matrix given in \eqref{invmetrics} is different from the $\eta'$ matrices previously used in \cite{KKV2, Sahay:2013xda}. Equivalently, the embedding of 
$\mathrm{K} = \mathrm{SO}(2,2)\times\mathrm{SO}(2,2)$ inside $\mathrm{G} = \mathrm{SO}(4,4)$ is different compared to those references. This is because in our work dimensional reduction is performed differently: we first do a timelike reduction from 6d to 5d, and then do two spacelike reductions from 5d to 3d. The details can be found in appendix~\ref{app:redux}.

All propagating degrees of freedom of the STU model reduced to three dimensions can be written as scalar fields and are summarised by an element $V\in \mathrm{G}/\mathrm{K}$. 
The dynamics are constructed from the $\mathfrak{so}(4,4)$ Lie algebra element $\partial_\mu V \cdot V^{-1}$ that is decomposed into
\begin{align}
P_\mu = \frac12\left(\partial_\mu V\cdot V^{-1} + ( \partial_\mu V\cdot V^{-1})^\sharp \right),\qquad Q_\mu = \frac12\left(\partial_\mu V\cdot V^{-1} - ( \partial_\mu V\cdot V^{-1})^\sharp \right).
\end{align}
The bosonic dynamics of STU supergravity reduced to three dimensions then is given by
\begin{align}
\label{3dL}
\mathcal{L}_{(3)} = \sqrt{g} \left( R -\frac12 g^{\mu\nu}\mathrm{Tr}(P_\mu P_\nu) \right),
\end{align}
where $g_{\mu\nu}$ is the (non-propagating) three-dimensional metric. This is an instance of a general $\mathrm{G}/\mathrm{K}$ $\sigma$-model that was discussed 
for example in~\cite{Breitenlohner:1987dg} and in the context of inverse scattering in~\cite{KKV}.

The action of $\mathrm{G}$ on $V$ is by a local $k(x)\in \mathrm{K}$ and a global $g\in \mathrm{G}$ via
\begin{align}
V(x) \to k(x) V(x) g.
\end{align}
It is also useful to define the function $M$  
\begin{align}
\label{sM}
M(x) = V^\sharp (x) V (x)\quad\quad
\textrm{with}\quad
M(x) \to g^\sharp M(x) g,
\end{align}
which is often easier to work with since it avoids the local $k$-transformation. It obeys $M^\sharp = M$.

When system (\ref{3dL}) is further reduced to two dimensions, the equations of motion become completely integrable and the group of symmetry transformations in the space of solutions
is infinite-dimensional. The latter is called the Geroch group and it is defined as the affine extension of the group $\mathrm{G}$~\cite{BM}. We now review this construction briefly and refer 
the reader to~\cite{KKV,KKV2} for a more detailed account. 

In a suitable coordinate system, the metric can be written in a canonical form with all functions depending on two variables. The two-dimensional base metric has the form
\begin{align}
ds_{2}^2=f^2 (d\rho^2 +dz^2),
\end{align}
where $(\rho,z)$ are the so-called Weyl canonical coordinates and the function $f(\rho,z)$ is referred to as the  conformal factor. Using the coordinates  $ x^{\pm}=\frac{1}{2}(z\mp i \rho)$, the equations of motion read
\begin{subequations}
\begin{align}
 \pm if^{-1}\partial_{\pm}f=\frac{\rho}{4}\mathrm{Tr}\left(P_{\pm}P_{\pm}\right),\label{feqn}\\
 D_m\left(\rho P^{m}\right)=0\label{seqn},
\end{align}
\end{subequations}
where $D_m P_n=\partial_m P_n -[Q_m,P_n]$. The first equation for the conformal factor can be solved by simple integration when $P_{\pm}$ is known, so the main task is to solve equation (\ref{seqn}).
This is a non-linear equation which poses a problem that can be equivalently presented as a linear system of equations (Lax pair). In the latter
formulation, (\ref{seqn}) is viewed as the compatibility condition for the set of linear equations, which take the form
\begin{align}
\label{BMlinsys}
 \partial_\pm \cV(t,x) \cV(t,x)^{-1}=\frac{1\mp it}{1\pm it}P_\pm(x) +Q_\pm(x),
\end{align}
where $t$ is the spectral parameter known to appear in Lax pairs. The equivalence of~\eqref{BMlinsys} with~\eqref{seqn} requires that the spectral parameter $t$ is a function of $x$ according to
\begin{align}
t_\pm=\frac{1}{\rho}\left[(z-w)\pm\sqrt{(z-w)^2 +\rho^2}\right].
\end{align}
In the above expression, $w$ is an integration constant that is later used as an $x$-independent spectral parameter. In the following, we will mean $t$ to correspond to the $t_+$ solution.

The generating function $\cV(t,x)$ is the generalised coset element that satisfies:
\begin{align}
\label{limV}
 \lim_{t \to 0}\cV(t)=V.
\end{align}
Similarly to the finite group elements, a more useful object to work with is the ``monodromy matrix'' $\cM$ defined as
\begin{align}
\cM(w)=\left(\cV(t,x)\right)^{\sharp}\cV(t,x)\quad\Longrightarrow \quad \left(\cM(w)\right)^\sharp =\cM(w),
\end{align}
where the generalisation of the $\sharp$ operation to $t$-dependent matrices is given by
\begin{align}
 \left(\cV(t)\right)^{\sharp}=\cV^{\sharp}\left(-\frac{1}{t}\right).
\end{align}
Under $w$-dependent global transformations, the monodromy matrix transforms as $\cM(w)\rightarrow \cM^{g}(w):=g^{\sharp}(w)\cM(w)g(w)$. The linear system~\eqref{BMlinsys} implies that $\cM(w)$ is constant.

In order to find solutions of the STU model, we use an inverse scattering technique based on the linear system~\eqref{BMlinsys} and restrict to the soliton sector~\cite{BMnotes,BV,KKV,KKV2}. In the soliton sector, one assumes a simple meromorphic form of $\cM(w)$ and the method amounts to  a series of purely algebraic steps that allows to find the space-time solution.

For STU supergravity and solutions that are asymptotically flat in $D=5$, we start with the following ansatz for the monodromy matrix 
\begin{subequations}
\bea
\label{cMansatz}
\cM(w) &=& Y + \sum_{k=1}^N \frac{A_k}{w-w_k}, \\
\cM^{-1}(w) &=& \h \cM^{T}\h=\h\left(Y + \sum_{k=1}^N \frac{A^{T}_k}{w-w_k}\right)\h,
\eea
\end{subequations}
where $w_k$ are the pole locations and $\h$ is the $\mathrm{SO}(4,4)$-invariant metric from~\eqref{invmetrics}. The matrix $Y$ is a constant matrix\footnote{In \cite{KKV2} $Y$ is specified to the unit matrix, since we were working on four dimensional
asymptotically flat solutions. The case of five-dimensional asymptotically flat solutions requires changes in the form of $Y$ that we will derive below.} 
 such that $\cM(\infty)=Y$. The residues satisfy $A_k^\sharp=A_k$.
 
 Our aim is to factorise $\cM(w)$ as
\be
\label{Afactors}
\cM(w)=A_{-}^{\sharp}(t,x)M(x)A_{+}(t,x),
\ee
where the matrices $A_{+}(t,x)$ and $A_{-}(t,x)=A_+(-1/t,x)$  are in $\mathrm{SO}(4,4)$ and satisfy
\be
A_{+}(0,x)=\id=A_{-}(\infty,x).
\ee
The matrix $M(x)$ is the sought-after solution, defined in (\ref{sM}), from which the physical fields in $V$ can be read off.

To start the factorisation procedure, we write $\cM(w)$ as a function of $(t,x)$ using the relations
\be
  \frac{1}{w-w_k} = \nu_k \left( \frac{t_k}{t-t_k}+ \frac{1}{1+t t_k}\right)
 \ee
  and
\be
\nu_k = -\frac{2}{\rho\left(t_k + \frac1{t_k}\right)},
\ee
with $t_k=\frac{1}{\rho}\left((z-w_k)+\sqrt{(z-w_k)^{2}+\rho^2}\right)$.
We arrive at
\begin{align}
\label{texpansion}
 \cM(t,x)=Y+\sum_{k=1}^N \frac{\nu_k t_k A_k}{t-t_k}+\sum_{k=1}^N \frac{\nu_k A_k}{1+tt_k}\,,
\end{align}
where the rank-2 residue matrices $A_k$ are factorized in terms of 8-dimensional constant vectors $a_k\,,b_k$ as follows
\begin{align}
A_k =  \alpha_k a_k a_k^T \h'-\beta_k (\h b_k)(\h b_k)^{T}\h',
\label{Ak}
\end{align}
with $\alpha_k\,,\beta_k$ constant parameters and $\h'$ the metric preserved by $\mathrm{SO}(2,2)\times \mathrm{SO}(2,2)$. This form and the rank condition follow from 
an analysis of the condition $A_k^\sharp=A_k$ and from the embedding of pure gravity solutions into STU~\cite{KKV2}.

The vectors $a_k\,,b_k$ are required to satisfy the conditions
\begin{subequations}
\label{vecconds}
 \bea
\label{aconds}
 a^{T}_k\h a_k&=&0,\\
 \label{bconds}
 b^{T}_k\h b_k&=&0,\\
 \label{aborthog}
 a^{T}_k b_k&=&0,
\eea
\end{subequations}
for all $k$, which stem from the requirement that the product $\cM(w)\cM(w)^{-1}$ have no double poles.
Next, we need to determine the matrix $A_{+}$ that is of the form
\be
   A_{+}(t)=\id-\sum_{k=1}^N \frac{t C_k}{1+tt_k}\,,
\ee
where $C_k$ are matrices parametrized as 
\be
 C_k=c_k a^{T}_k \h'-(\h d_k) (\h b_k)^{T}\h'\,.
\ee
 The vectors $c_k\,,d_k$ are obtained from the matrix equations 
 \begin{subequations}
\bea
 \label{cmat}
 c &=& \h' b \Gamma^{-1},\\
\label{dmat}
 d &=& \h' a \left(\Gamma^{T}\right)^{-1}\,,
 \eea
\end{subequations}
where $a, b, c$, and $d$ are $8\times N$ matrices with columns the vectors $a_k, b_k, c_k$, $d_k$ respectively. The  $N\times N$ matrix $\Gamma$  with elements
\begin{align}
 \label{Gdef}
\Gamma_{kl} = \left\{ \begin{array}{ll}
 \frac{\gamma_k}{t_k} &\mbox{\qquad for \qquad $k=l$} \\
  \frac{a_k^T b_l}{t_k-t_l}  &\mbox{\qquad for \qquad $k \neq l$}
       \end{array} \right.
\end{align}
is acquired by solving the following equations\footnote{We arrive at these relations starting from the conditions for no single poles in the product $\cM(t,x)\cM(t,x)^{-1}$.} for the numbers $\gamma_k$
\begin{subequations}
\bea
\left[\left.\left(\cM(t,x)-\frac{\nu_k A_k}{1+tt_k}\right)\right |_{t \rightarrow -\frac{1}{t_k}}\right]\h \h' a_k &=& \nu_k \beta_k \gamma_k (\h b_k), \label{gammaone} \\
(\h b_k)^{T}\h' \h \left[\left.\left(\cM(t,x)-\frac{\nu_k A_k}{1+tt_k}\right)\right |_{t \rightarrow -\frac{1}{t_k}}\right]^{T} &=& \nu_k \alpha_k \gamma_k a^{T}_k.  \label{gammatwo}
\eea
\end{subequations}
The final step in the process is to take the limit  $t\rightarrow \infty$ of (\ref{Afactors}) and thus find the new solution $M(x)$ :
\begin{align}
\label{sMfinal}
 M(x)=Y A_{+}^{-1}(\infty).
\end{align}
The full solution to the equations of the theory is constructed once the conformal factor is determined:
\be
\label{CF}
f^2 = k_{\rom{BM}} \cdot \prod_{k=1}^{N} (t_k \nu_k) \cdot \det \Gamma,
\ee
where $k_{\rom{BM}}$ is an integration constant.
The detailed calculations leading to the final formula outlined above can be found in \cite{KKV}, \cite{KKV2}.

\subsection*{Asymptotic behavior of $M(x)$, $\cM(w)$}
Let us start with five-dimensional Minkowski space that is trivially uplifted to six dimensions along the $y$-direction
\be
ds^2=-dt^2+dy^2+dr^2+r^2\left[d\theta^2+\sin^2 \theta d\phi^2+\cos^2\theta d\psi^2\right],
\ee
where $\theta \in [0,\frac{\pi}{2}]$ , $\phi,\psi$ are standard angular coordinates with range $[0,2\pi)$ and $y$ is a periodic coordinate around a circle. Following \cite{Giusto:2007fx},
we change to the coordinates\footnote{The specific normalization for these coordinates is chosen to simplify later expressions.}
\be
\label{GScoords}
\phi_{+}=\frac{1}{2}(\psi+\phi),\quad \phi_{-}=(\phi-\psi)
\ee
and obtain the metric
\be
ds^2=-dt^2+dy^2+dr^2+r^2\left[d\theta^2+\frac{1}{4}d\phi_{-}^2+d\phi_{+}^2 - \cos 2 \theta d\phi_{-}d\phi_{+}\right].
\ee
As will become clear shortly, the advantage of changing coordinates in this way is that the coset matrices asymptotically tend to constant values (see discussion in \cite{Giusto:2007fx},\cite{CV_Geroch}).
Had we left the standard angular coordinates, we  would encounter infinities in the asymptotic behavior of our solution, which would in turn require us to include poles at infinity in the ansatz (\ref{cMansatz}).
As it is not yet clear how to incorporate this kind of poles in the formalism presented here, we choose to work with the ``nicer'' coordinates (\ref{GScoords}).
\par Upon dimensional reduction along the directions $t,\phi_{+}$ and $y$, the above metric corresponds to the following expressions for the fields in three dimensions
(details on the structure of the three-dimensional theory
are found in section \ref{subsec:D=3}):
\bea
e^{2U} &=& r,\quad y^1=y^2=y^3=r,\quad \tilde{\zeta_0}=r^2, \\
\quad A^{0} &=& -\frac{1}{2} \cos 2 \theta d\phi_{-}\,, \\
ds_3^2 &=& r^2 \left[dr^2 + r^2 d \theta^2 + r^2 \cos^2 \theta \sin^2 \theta d \phi_-^2 \right], \label{3dbase5dflat}  
\eea
while the rest of the fields vanish. At this point it is important to note that, as opposed to flat space in four dimensions, the 3d scalar fields and one-forms have a non-trivial profile. The 
 matrix $M(x)$ reads
\bea
M(x) = \left(
\begin{array}{cccccccc}
 \frac{1}{r^2} & 0 & 0 & 0 & 0 & 0 & -1 & 0 \\
 0 & 1 & 0 & 0 & 0 & 0 & 0 & 0 \\
 0 & 0 &  \frac{1}{r^2} & 0 & 1 & 0 & 0 & 0 \\
 0 & 0 & 0 & 1 & 0 & 0 & 0 & 0 \\
 0 & 0 & -1 & 0 & 0 & 0 & 0 & 0 \\
 0 & 0 & 0 & 0 & 0 & 1 & 0 & 0 \\
 1 & 0 & 0 & 0 & 0 & 0 & 0 & 0 \\
 0 & 0 & 0 & 0 & 0 & 0 & 0 & 1 \\
\end{array}
\right),
\label{sMflat}
\eea
which in the limit $r\rightarrow \infty$ takes on the constant value
\bea
Y = \left(
\begin{array}{cccccccc}
 0 & 0 & 0 & 0 & 0 & 0 & -1 & 0 \\
 0 & 1 & 0 & 0 & 0 & 0 & 0 & 0 \\
 0 & 0 & 0 & 0 & 1 & 0 & 0 & 0 \\
 0 & 0 & 0 & 1 & 0 & 0 & 0 & 0 \\
 0 & 0 & -1 & 0 & 0 & 0 & 0 & 0 \\
 0 & 0 & 0 & 0 & 0 & 1 & 0 & 0 \\
 1 & 0 & 0 & 0 & 0 & 0 & 0 & 0 \\
 0 & 0 & 0 & 0 & 0 & 0 & 0 & 1 \\
\end{array}
\right).
\label{Ymatrix}
\eea
For solutions that are asymptotically flat in five dimensions, we therefore require that the monodromy matrix $\cM(w)$ asymptotes to $Y$ as $w\to\infty$ as shown in the ansatz~\eqref{cMansatz}. We note that $Y^\sharp=Y$.

\subsection*{Charging transformations}

In order to apply a charging transformation to a seed solution $\cM(w)$, we need to find the subgroup of elements that preserve the asymptotic behavior determined by $Y$ above. Thus we are looking for elements $g_D\in \mathrm{SO}(4,4)$ such that 
\begin{align}
\label{Dcond}
g_D^\sharp Y g_D=Y.
\end{align}
{}From the observation that there is an $SO(4,4)$, ``$\sharp$''-invariant matrix that satisfies
\be
D^\sharp  D = Y
\ee
with
\bea
D &=& \frac{1}{\sqrt{2}}\left(
\begin{array}{cccccccc}
 1 & 0 & 0 & 0 & 0 & 0 & -1 & 0 \\
 0 & -\sqrt{2} & 0 & 0 & 0 & 0 & 0 & 0 \\
 0 & 0 & 1 & 0 & 1 & 0 & 0 & 0 \\
 0 & 0 & 0 & \sqrt{2} & 0 & 0 & 0 & 0 \\
 0 & 0 & -1 & 0 & 1 & 0 & 0 & 0 \\
 0 & 0 & 0 & 0 & 0 & -\sqrt{2} & 0 & 0 \\
 1 & 0 & 0 & 0 & 0 & 0 & 1 & 0 \\
 0 & 0 & 0 & 0 & 0 & 0 & 0 & \sqrt{2} \\
\end{array}
\right),
\label{Dmatrix}
\eea
we deduce that the appropriate charging element must be of the form
\be
\label{gD}
g_D=D^{-1} k D,\qquad k \in \mathrm{K}=\mathrm{SO}(2,2)\times\mathrm{SO}(2,2),
\ee
where $\mathrm{K}$ is the subgroup defined in~\eqref{GK}. Therefore the solutions to~\eqref{Dcond} form an $\mathrm{SO}(2,2)\times\mathrm{SO}(2,2)$ subgroup conjugate to $\mathrm{K}$.
Their action on the monodromy matrix
\be
\cM^{g_D}(w):=g_{D}^{\sharp}(w)\cM(w)g_{D}(w)
\ee 
preserves the form (\ref{cMansatz}) and therefore the five-dimensional asymptotics. See also \cite{Giusto:2007fx, CV_Geroch} for a similar discussion in the $\mathrm{SL}(3,\mathbb{R})/\mathrm{SO}(2,1)$ case.

\section{Supergravity configuration}
\label{sec:sugra}
Having reviewed the inverse scattering formalism for STU supergravity we now present the construction of the relevant supergravity configuration. In the following section we analyse the resulting 
configuration and relate it to the JMaRT fuzzball. 
The steps we follow are:
\begin{enumerate}
\item We first construct an appropriate Euclidean five-dimensional gravity configuration trivially lifted to six-dimensions along the time direction, i.e., a metric of the form
\bea
ds^2_6 = -dt^2 + ds^2_5.
\eea
\item On this configuration we apply an appropriate SO(4,4) charging transformation 
\bea
\cM_\rom{new}(w) = g^\sharp_D \cM_\rom{old}(w) g_D,
\eea
with an appropriate $g_D$ to add electric charges. $g_D$ is of the form~\eqref{gD} in order to preserve the five-dimensional asymptotics.
\item Then we analyse degeneration properties of various Killing vectors and relate the final configuration to the JMaRT fuzzball. 
\end{enumerate}

In the above three-step process the second and third steps are fairly standard. The first step however requires some explanation. 
As is well known in the inverse scattering literature \cite{ER, BV}, to obtain a single center non-extremal black hole, e.g., Kerr, a two soliton transformation is required on an 
appropriate seed solution. The seed solution in the case of Kerr is simply the four-dimensional Minkowski space. The pole locations for such a transformation can be taken to be real or complex. 
In the case of transformation with 
complex conjugate poles one obtains the Kerr solution that is 
``over-rotating'', i.e., a solution without horizons and with
a naked singularity. On the other hand, if both poles are taken to be real, the
solution obtained is  ``under-rotating'', i.e., with curvature singularity behind an event horizon~\cite{BV}.

The JMaRT fuzzball \cite{JMaRT} was obtained by studying certain limits of the over-rotating five-dimensional Cveti{\v c}--Youm metrics. Therefore, it seems that in order to construct 
the JMaRT fuzzball by an inverse scattering method one must first construct the over-rotating Myers-Perry metric and then by adding appropriate charges construct the over-rotating Cveti{\v c}--Youm metric. 
In analogy with the Kerr example discussed in the previous paragraph this procedure  would seem to require working with complex conjugate poles. This is undesirable:
the inverse scattering formalism reviewed in the previous section is adapted to real poles and the conditions on the residue matrices and vectors would need to be adapted in order to ensure that the resulting monodromy matrix lies in $\mathrm{SO}(4,4)$. We therefore choose a different approach.

Our main observation that bypasses this difficulty is the following: the statement that in the inverse scattering construction the
obtained solution is under-rotating ---when both poles are taken to be real--- is a statement in the context of \textit{Lorentzian} four-dimensional vacuum gravity. In the context of \textit{Euclidean} four-dimensional vacuum gravity ---when both poles are 
taken to be real--- the obtained solution turns out to be the Kerr instanton. We find that the same  picture applies in our construction. Via a two-soliton ansatz in the STU 
set-up we construct a Euclidean five-dimensional vacuum gravity configuration. This object turns out to be the Myers--Perry instanton. We trivially lift this object 
to six-dimensions along the time direction.  We apply an appropriate $\mathrm{SO}(4,4)$ charging transformation to add electric charges. The resulting configuration is shown to be related to the JMaRT fuzzball.

We start with an $\mathrm{SO}(4,4)$ monodromy matrix of the form (cf.~\eqref{cMansatz})
\be
\label{M11}
\cM(w) = Y + \frac{A_1}{w-c} + \frac{A_2}{w+c},
\ee
where the residue matrices $A_1$ and $A_2$ are parameterized as
\begin{subequations}
\label{Ak2}
\begin{align}
A_1 &=  \alpha_1 a_1 a_1^T \h'-\beta_1 (\h b_1)(\h b_1)^{T}\h', \\
A_2 &=  \alpha_2 a_2 a_2^T \h'-\beta_2 (\h b_2)(\h b_2)^{T}\h',
\end{align}
\end{subequations}
and the $Y$ matrix was introduced in the previous section.
The monodromy matrix has two real poles at locations $w = \pm c$.
For the Euclidean five-dimensional vacuum gravity configuration we are interested in, we choose the vectors to be of the form
\begin{subequations}
\begin{align}
a_1 &= \{1,0,0,\zeta_{12},0,0,\zeta_{11},0\}, \\
a_2 &= \{\zeta_{21},0,0,1,0,0,\zeta_{22},0\}.
\end{align}
\end{subequations}
This form of the vectors can be easily guessed, for example, by examining the general form of the matrix $M(x)$ in the truncation of interest.

Next we introduce the notation $a = (a_1 \ \ a_2)$ where the vectors $a_1$ and $a_2$ are put as column vectors in a $8 \times 2$ matrix $a$ \cite{BMnotes}. Then we construct a $2 \times 2$ matrix $\xi$
\be
\xi  = a^T \eta' Y^{-1} a =
\left(
\begin{array}{cc}
 a_1^T \eta' Y^{-1}a_1 &  a_1^T \eta' Y^{-1}a_2 \\
  a_2^T \eta' Y^{-1}a_1 &  a_2^T \eta' Y^{-1}a_2
\end{array}
\right),
\label{xi}
\ee
where the matrices $Y$ and $\h'$ are defined in equations \eqref{Ymatrix} and \eqref{invmetrics} respectively. We also note that
\be
(\eta' Y^{-1})^T = \eta' Y^{-1},
\ee
as a result the $\xi$ matrix is symmetric. Using the $\xi$ matrix we assign 
\begin{subequations}
\begin{align}
\alpha_1 &= \frac{2c}{\det \xi} \xi_{22}, & \alpha_2 &= -\frac{2c}{\det \xi} \xi_{11}, \\
\beta_1 &= - \frac{1}{\det \xi} \alpha_1,  & \beta_2 &= - \frac{1}{\det \xi} \alpha_2, &
\end{align}
\end{subequations}
and choose $b$-vectors as
\begin{align}
b &= (\det \xi) \eta' Y^{-1} a  \xi^{-1} \epsilon,&
\epsilon &=
\left(
\begin{array}{cc}
0 & -1 \\
1 & 0
\end{array}
\right).
\end{align}
The $a$ and $b$ vectors and the $\alpha$ and $\beta$ parameters obtained in this way satisfy all coset constraints from section~\ref{sec:prelim}. Moreover, 
\be
a^T b = (\det \xi) \epsilon.
\ee
Now following the factorization algorithm of the previous section, we obtain a spacetime configuration. 
For details on the SO(4,4) sigma model we refer the reader to appendix \ref{app:redux}.
For this configuration it turns out that the dilatonic scalars $y^I$ are all equal $(I=1,2,3)$, and 
$x^I, \zeta^I, \tilde \zeta_I$ are all zero,
\begin{subequations}
\begin{align}
y^I &= y, & \tilde \zeta_I & =  0, \\
\zeta^I & =  0, & x^I  & = 0.
\end{align}
\end{subequations}
This precisely corresponds to the truncation to the Euclidean five-dimensional vacuum sector, with the six-dimensional metric of the form
\be
ds^2_6 = - dt^2 + ds^2_5.
\ee
In terms of the entries of the matrix $M(x)\equiv m_{ab}$, among the remaining fields $(U,y, \zeta^0, \tilde \zeta_0)$ take values
\begin{subequations}
\label{scalarsFromM}
\begin{align}
y &= \sqrt{\frac{m_{44}}{m_{33}}}, &
e^{2U} &= \frac{1}{\sqrt{m_{44}m_{33}}},\\
\zeta^0 &= - \frac{m_{41}}{m_{44}}, &
\tilde \zeta_0 &=  \frac{m_{35}}{m_{33}}, 
\end{align}
\end{subequations}
and $\sigma$ takes value
\be
\label{eq:sig}
\sigma = -\frac{m_{35}m_{41} + 2 m_{33} m_{47}}{m_{33}m_{44}}.
\ee
In this truncation various matrices take simple forms, for example, the imaginary part of the matrix $\cal{N}$ occurring in the reconstruction of the higher-dimensional solution (cf. appendix~\ref{app:red4}) is simply
\bea
\mathrm{Im} \, \mathcal{N} = \left(
\begin{array}{cccc}
 -1 & 0 & 0 & 0 \\
 0 & \frac{m_{44}}{m_{33}} & 0 & 0 \\
 0 & 0 & \frac{m_{44}}{m_{33}} & 0 \\
 0 & 0 & 0 & \frac{m_{44}}{m_{33}}
\end{array}
\right).
\eea
At this stage we set 
\bea
\zeta_{12} &= 0, \quad
\zeta_{21} &= 0. 
\eea
This is a simplification we do in order to make the computations simpler and the presentation more transparent. In this work we are interested in the singly rotating fuzzball. The parameters $\zeta_{11}$ and 
$\zeta_{22}$ are sufficient to parameterize the singly rotating configuration. With $\zeta_{12} \neq \zeta_{21} \neq 0$ one should be able to construct the doubly rotating JMaRT fuzzball, but we do not attempt this here.

With these simplifications, we get the one-form $\omega_3$ to be 
\be
\omega_3 = -\frac{2 c \zeta_{22} \left(u^2 \left(\zeta_{22}^2-2
   \zeta_{11}\right)+2 \zeta_{11} u \left(v^2-1\right)+v^2
   \left(2 \zeta_{11}-\zeta_{22}^2\right)\right)}{\left(2
   \zeta_{11}-\zeta_{22}^2\right) \left(2 \zeta_{11}
   \left(u^2-1\right)+\zeta_{22}^2 \left(v^2-u^2\right)\right)} dz_3,
\ee
and the rest of the three-dimensional one-forms all vanish. In writing this equation we have introduced prolate spherical coordinates $(u,v)$  defined via the relations
\be
\rho = 4 c \frac{t_1 t_2}{(t_2-t_1)(t_1 t_2 + 1)}, \qquad \qquad z = \frac{c(t_1+t_2)(t_1 t_2 - 1)}{(t_2 -t_1)(t_1 t_2+1)},
\ee
and
\be
t_1 = \frac{(u-1)(1+v)}{\sqrt{(u^2-1)(1-v^2)}}, \qquad \qquad t_2 = \frac{(u+1)(1+v)}{\sqrt{(u^2-1)(1-v^2)}},
\label{t1t2prolate}
\ee
with $u \in [1, \infty)$ and $v \in [-1,1]$.
The above relations are most useful in doing the computations. After the factorization of $\cM(w)$, the matrix $M(x)$ is naturally written in terms of variables $t_1$ and $t_2$ which are exactly the values of the $t$-spectral parameter at the pole positions as discussed in section~\ref{sec:prelim}. It is computationally 
most efficient to convert these variables in terms of the prolate spherical coordinates via \eqref{t1t2prolate}. Another set of expressions for going from canonical coordinates to prolate coordinates is,
\begin{align}
\rho &= c \sqrt{(u^2-1)(1-v^2)}, & z &= c u v.
\end{align}
The three-dimensional base metric in the same set of coordinates is 
\bea
ds_3^2 &=& 
 \frac{\left(\zeta_{22}^2-2 \zeta_{11}\right)}{16}
   \left((u^2-v^2)\zeta_{22}^2 -2 \left(u^2-1\right) \zeta_{11}\right) 
\left[
\frac{du^2}{\left(u^2-1\right)} +
\frac{dv^2}{\left(1-v^2\right)}  
\right] \nn \\
& &
+
c^2 \left(u^2-1\right) \left(1-v^2\right) dz_3^2,
\label{basemetric}
\eea
where the integration constant $k_\rom{BM}$ for the conformal factor has been fixed by the requirement of asymptotic flatness. The values of the matrix $M(x)$ needed to construct the scalar fields according to~\eqref{scalarsFromM} are slightly unwieldy and are given in appendix~\ref{app:Mvalues}.

This completes the first step.  On this configuration we act with the following SO(4,4) group element
\be
g_D = D^{-1} g D,
\ee
as
\be
M_\rom{new} = g_D^\sharp M_\rom{old} g_D, \label{groupaction}
\ee
where $g$ is
\be
g = \exp \left[ \left( i \frac{\pi}{2}- \delta_2 \right) K_{q_2} \right] \cdot \exp \left[ \left( i \frac{\pi}{2}- \delta_3 \right) K_{q_3} \right]
\label{groupelement}
\ee
and $K_{q_I}=E_{q_I}+F_{q_I}$ are combinations of the K generators that are described in appendix~\ref{subsec:D=3}.
The matrix $D$ is defined in equation \eqref{Dmatrix}.
This group element adds two charges parameterized by $\delta_2$ and $\delta_3$. To avoid notational clutter we use the shorthand $c_{2,3} =\cosh  \delta_{2,3}$ and $s_{2,3} =\sinh  \delta_{2,3}$.  This 
group 
element preserves the asymptotic matrix $Y$ by construction. The $i \frac{\pi}{2}$ shifts with the generators $K_{q_2}$ and $K_{q_3}$ are not necessary. However,
they are very convenient, as these shifts result in a parameterization of the final solution that directly matches with the presentation of the JMaRT paper \cite{JMaRT}. A discussion of what these shifts
 correspond to from the coset model perspective can be found in section \ref{shifts} of the appendix. We note that $g$ lies in $\mathrm{K}$ despite the $i\frac{\pi}{2}$. 

{\allowdisplaybreaks 
Since the group action \eqref{groupaction} is a global SO(4,4) rotation it does not change the base metric \eqref{basemetric}. It changes the rest of the fields. 
The final expressions for the resulting scalars 
are somewhat cumbersome. All sixteen scalars that specify the configuration can be read from the matrix 
$M_\rom{new}$. Five of these scalars namely $\sigma$ and $\tilde \zeta_\Lambda$ need to be dualized in order to find the dual one-forms. The intermediate expressions are not particularly illuminating\footnote{Mathematica files with details are available upon request to the authors.}, 
we 
only present a final set of expressions for the Killing part of the six-dimensional metric:
\be
G_\rom{Killing} =  \left(\sqrt{\zeta_{11} \left(4 c_2^2-2 (u+1)\right)+\zeta_{22}^2 (u+v)} \sqrt{\zeta_{11} \left(4 c_3^2-2
   (u+1)\right)+\zeta_{22}^2 (u+v)}\right)^{-1} g
\ee
with the $4\times 4$ matrix $g$ having entries 
\begin{subequations}
\bea
g_{33} &=& \frac{1}{2
   \zeta_{11}-\zeta_{22}^2} \Big{[}2 c^2 \left(\zeta_{22}^2 (u+v) \left(2 s_2^2+2
   s_3^2-u+v+2\right) \right. \nn \\
& & \qquad \qquad \ \ \left. +2 \zeta_{11} \left(2
   s_2^2-u+1\right) \left(2 s_3^2-u+1\right)\right)\Big{]},\\
g_{34} &=& 2 c \zeta_{22} s_2 s_3 (u+v), \\
g_{35} &=& 2 c \zeta_{11} v \left(2 s_2^2-u+1\right) \left(-2
   s_3^2+u-1\right)+c \zeta_{22}^2 (u+v) (u v-1),\\
g_{36} &=& 2 c \zeta_{22} c_2 c_3 (u+v),\\
g_{44} &=& \zeta_{22}^2 (u+v)-2 \zeta_{11} (u-1),\\
g_{45} &=&-2 \zeta_{11} \zeta_{22} s_2 s_3 (v+1),\\
g_{46} &=& 0,\\
g_{55} &=&\zeta_{11} \zeta_{22}^2 \left(2 u
   \left(s_2^2+s_3^2+1\right)-2 v
   \left(s_2^2+s_3^2+1\right)-2 u^2+v^2+1\right)\nn \\ & & -2 \zeta_{11}^2 \left(2 s_2^2-u+1\right) \left(-2
   s_3^2+u-1\right)+\frac{1}{2} \zeta_{22}^4 (u-v) (u+v), \\
g_{56} &=& 2 \zeta_{11} \zeta_{22} c_2 c_3 (v-1),\\
g_{66}  &=& 2 \zeta_{11} (u+1)-\zeta_{22}^2 (u+v).
\eea
\end{subequations}
The three-dimensional one-forms obtained by dualisation of the scalar fields $\sigma$ and $\tilde \zeta_\Lambda$ can be found for completeness in appendix~\ref{app:dualone}.}

\section{Rod-structure analysis and the JMaRT fuzzball}
\label{sec:rods}

To verify that the above fields describe the JMaRT fuzzball we look for degeneration properties of the various Killing vectors. 
For this analysis we make use of the rod diagram representations from \cite{Harmark:2004rm}. Generically, as in the Belinski--Zakharov method, the configuration obtained after 
the above inverse scattering procedure does not have any standard orientation for its rods, i.e., for the 
five-dimensional asymptotically flat configuration the semi-infinite rods generically do not coincide with the $\phi$ and $\psi$ directions. This situation can however be 
remedied by making a linear coordinate transformation,
\be
G_\rom{final} = \Lambda^T G \Lambda,
\ee
where $G$ denotes the $4 \times 4$ Killing part of the metric given above and  $\Lambda \in \mathrm{SL}(4, \RR)$. The role of matrix $\Lambda$ is to make manifest the desired asymptotic behaviour. 
The following choice
\bea
\Lambda =\left(
\begin{array}{cccc}
\frac{\left(\zeta_{22}^2-2 \zeta_{11}\right)}{4 c} &
   0 & -\frac{\left(\zeta_{22}^2-2 \text{$\zeta
   $11}\right)}{4 c} & 0 \\
- \frac{1}{2} s_2 s_3 \zeta_{22} & 1 & \frac{1}{2} s_2
   s_3 \zeta_{22} & 0 \\
 \frac{1}{2} & 0 & \frac{1}{2} & 0 \\
 \frac{1}{2}c_2 c_3 \zeta_{22} & 0 & -\frac{1}{2} c_2 c_3
    \zeta_{22} & 1
\end{array}
\right),   
\qquad 
\mbox{with action}
\qquad  
 \left(
\begin{array}{c}
z_3 \\
z_4 \\
z_5 \\
z_6
\end{array}
\right)
 =
\Lambda \left(
\begin{array}{c}
\phi \\
y \\
\psi \\
t
\end{array}
\right),
\eea
does the required job for us. The requirement that the $z_5$ and $z_3$ coordinates used above are  asymptotically
 $\phi_+ = \frac{1}{2}(\phi + \psi)$ and $\phi_- = \phi - \psi$  imposes the relation $c= \frac{1}{4}\left(\zeta_{22}^2 -2 \zeta_{11}\right)$. The rod structure of the resulting configuration is:
\begin{itemize}
 \item The semi-infinite rod $z \in (-\infty, -c]$ has orientation $\partial_\phi$.
\item The middle rod $z\in [-c,c]$ has orientation $
\partial_y + \frac{\zeta_{22}}{2\zeta_{11}s_2 s_3} \partial_\phi.$
\item The semi-infinite rod $z \in [+c, \infty)$ has orientation $\partial_\psi$. 
\end{itemize}
This is the structure of the JMaRT solution. To see that the configuration is precisely the JMaRT fuzzball, we write the metric in the standard radial and polar coordinate and compare 
it with the presentation of \cite{JMaRT}. 
First we change the parameterization
\bea
\zeta_{22} = -a_1, \quad
\zeta_{11} = \frac{1}{2} M.
\eea
The standard radial coordinate $r$ and polar coordinate $\theta$ are related to prolate coordinates as
\be
u = \frac{2 r^2}{a_1^2 - M} + 1, \qquad \qquad v = - \cos 2\theta,
\ee
equivalently
\be
r^2 = \frac{1}{2} (a_1^2 -M) (u-1),\qquad \qquad \cos^2\theta = \frac{1}{2}(1-v).
\ee
We obtain
\bea
ds^2_6 &=& \frac{1}{\sqrt{\tilde H_2 \tilde H_3}} \left[ -(f-M)(dt - (f-M)^{-1} M c_2 c_3 a_1 \cos^2\theta d \psi)^2 \right. \nn \\
&& \qquad \quad \left. + f (dy + f^{-1} M s_2 s_3 a_1 \sin^2\theta d\phi)^2\right] \nn \\
& &
+{\sqrt{\tilde H_2 \tilde H_3}} \left(\frac{dr^2}{r^2 + a_1^2 -M} + d\theta^2 + \frac{r^2 \sin^2\theta}{f} d\phi^2 + \frac{(r^2 + a_1^2 -M)\cos^2\theta}{f-M} d\psi^2\right),
\label{JMaRTFinal}
\eea
with
\begin{subequations}
\begin{align}
f&=r^2 + a_1^2 \sin^2\theta, \\
\tilde H_2 &= f + M \sinh^2 \delta_2,\\
\tilde H_3 &= f + M \sinh^2 \delta_3.
\end{align}
\end{subequations}
These are precisely the coordinates and parameters used in \cite{JMaRT}. The six-dimensional dilaton and the 2-form field also match exactly. The smoothness analysis applies exactly as in  \cite{JMaRT}. 

In order to better understand the relation with the Myers--Perry instanton we end this section with  some comments about the zero-charge limit. When the charge parameters $\delta_2$ and $\delta_3$ go to zero 
in the parameterization \eqref{JMaRTFinal}
we obtain the over-rotating Myers--Perry metric lifted to six-dimensions
\bea
ds^2_6 &=&  dy^2 - \left(1- \frac{M}{f} \right) \left(dt - (f - M)^{-1}M a_1 \cos^2\theta d\psi\right)^2  \nn \\
& & 
+ f \left(\frac{dr^2}{r^2  + a_1^2 -M} + d\theta^2 \right)+ r^2 \sin^2 \theta d\phi^2   
+  \frac{f(r^2  + a_1^2 -M)  \cos^2\theta}{f-M} d\psi^2 .
\label{MPmetric}
\eea

If we \textit{do not} perform the  $i \pi/2$ shifts in the parameters $\delta_2$ and $\delta_3$ cf.~\eqref{groupelement}, then we get the same physical solution but in a  different 
parameterization. That parameterization can be obtained by doing the replacement
\begin{align}
\delta_2 & = i \frac{\pi}{2} - \tilde \delta_2, & \delta_3 &= i \frac{\pi}{2} - \tilde \delta_3,
\label{shifted}
\end{align}
in \eqref{JMaRTFinal}. We obtain
\begin{align}
ds^2_6 &= \frac{1}{\sqrt{\tilde H_2 \tilde H_3}} \left[ -(f-M)(dt + (f-M)^{-1} M \tilde s_2 \tilde s_3 a_1 \cos^2\theta d \psi)^2 \right. \nn \\
& \qquad \quad \left. + f (dy - f^{-1} M \tilde c_2 \tilde c_3 a_1 \sin^2\theta d\phi)^2\right] \nn \\
& \quad
+{\sqrt{\tilde H_2 \tilde H_3}} \left(\frac{dr^2}{r^2 + a_1^2 -M} + d\theta^2 + \frac{r^2 \sin^2\theta}{f} d\phi^2 + \frac{(r^2 + a_1^2 -M)\cos^2\theta}{f-M} d\psi^2\right),
\label{JMaRTFinal2}
\end{align}
where
\begin{subequations}
\begin{align}
\tilde H_2 &= f - M \cosh^2 \tilde \delta_2,\\
\tilde H_3 &= f - M \cosh^2 \tilde \delta_3.
\end{align}
\end{subequations}
When the charge parameters  $\tilde \delta_2$ and $\tilde\delta_3$ go to zero, we obtain 
\bea
ds^2_6 &=&  -dt^2 + f(f-M)^{-1} (dy - f^{-1} M a_1 \sin^2\theta d\phi)^2 
 + \frac{(f-M)r^2 \sin^2\theta}{f} d\phi^2 
\nn \\ & & 
+ (r^2 + a_1^2 -M)\cos^2\theta d\psi^2 + (f-M) \left(\frac{dr^2}{r^2 + a_1^2 -M} + d\theta^2\right),
\eea
To recognize this metric let us shift the radial coordinate as $r^2 \to \tilde r^2 = r^2 + a_1^2 -M$. We obtain
\bea
ds^2_6 &=&  -dt^2 + (\tilde f + M) \tilde f ^{-1} (dy - (\tilde f + M)^{-1} M a_1 \sin^2\theta d\phi)^2 
 + \frac{\tilde f (\tilde r^2 -a_1^2 + M) \sin^2\theta}{\tilde f + M} d\phi^2 
\nn \\ & & 
+ \tilde r^2 \cos^2\theta d\psi^2 + \tilde f \left(\frac{d\tilde r^2}{ \tilde r^2 - a_1^2 +M} + d\theta^2\right),
\eea
with
\be
\tilde f = \tilde r^2 - a_1^2 \cos^2\theta.
\ee
This is nothing but the Euclidean Myers--Perry instanton lifted to six-dimensions along the time direction. It is obtained by the following analytic continuation of the Myers--Perry metric \eqref{MPmetric}
\be
t \to i y, \quad y \to i t, \quad a_1 \to -i a_1, \quad M \to -M, \quad \phi \leftrightarrow \psi, \quad \theta \to \frac{\pi}{2} - \theta, \quad r^2\to \tilde{r}^2.
\ee

\section{Discussion}
\label{sec:concl}

To summarize, in this paper we have presented an inverse scattering construction of the JMaRT fuzzball. The key element in our construction is the fact that in the shifted parameterization 
\eqref{shifted} the JMaRT fuzzball is smoothly connected to the Euclidean Myers--Perry instanton when the charge parameters go to zero. The Euclidean Myers--Perry instanton can be rather 
straightforwardly constructed using the inverse scattering method with real poles in the context of Euclidean five-dimensional gravity. Since five-dimensional STU supergravity admits a lift to six-dimensions, the three-dimensional hidden symmetry group SO(4,4) of the STU theory contains the Ehlers SL(4,~$\RR$) group of the vacuum six-dimensional gravity as a subgroup. Identifying the appropriate SL(3,~$\RR$) corresponding to the Ehlers group of the Euclidean  five-dimensional gravity truncation of interest we have presented our construction.

Our construction opens up the possibilities of obtaining multi-center non-supersymmetric fuzzballs systematically. There are many ways in which our study can be extended. For reasons of computational 
complexity we have not attempted a construction of the doubly rotating fuzzball in this paper. In principle, this should be possible to do within the framework of this paper; however,  details are likely to be tedious. If the STU inverse scattering formalism can be slightly modified to allow for non-trivial seeds, or for a pole at infinity in the monodromy matrix, then we believe that 
computations will become much simpler. In that case we need not work with the twisted dimensional reduction introduced in \cite{Giusto:2007fx}, but rather proceed with dimensional reduction along the more natural angular coordinates. More broadly, 
given our construction, it seems that putting appropriate charges on the multi-center five-dimensional instanton metrics is the most promising direction to explore in regard to obtaining multi-center non-supersymmetric fuzzballs. A similar set of ideas have been explored in a recent paper \cite{Bossard:2014yta}, where instead of Euclidean five-dimensional gravity instantons the authors work with Euclidean Einstein-Maxwell instantons. 

A further exploration of this circle of ideas can lead to a general understanding of non-supersymmetric fuzzballs, which in view of Mathur's fuzzball proposal~\cite{Mathur:2005zp}
will help us understand better the nature of black hole entropy.

\subsubsection*{Acknowledgements} The authors are grateful to G. Bossard and J.~V.~Rocha for informative discussions. AK and AV gratefully acknowledge the hospitality of KITPC, Beijing, during the ``Quantum Gravity, Black Holes and Strings'' program where part of this work was done. AV would also like to thank the organizers of ``Black Objects Beyond Supersymmetry'' workshop at Utrecht University for warm hospitality.

\appendix
\section{Dimensional reduction}
\label{app:redux}

In this appendix, we give some more details on the construction of the scalar $\mathrm{SO}(4,4)$ $\sigma$-model that underlies the STU-fuzzball. For this we start with the well-known truncation of type IIB supergravity on $T^4$ to a consistent subsector with Lagrangian
\begin{align}
\mathcal{L}_{(6)} =  R -\frac12 (\partial\Phi)^2 - \frac1{12}e^{-\sqrt{2}\Phi} H_{MNP}H^{MNP},
\end{align}
where $H_{MNP}=3\partial_{[M} B_{NP]}$ is the field strength of the RR-two-form. We will perform the reduction of this theory from $D=6$ down to $D=3$ in the following order
\begin{align}
D=6 \stackrel{t}{\longrightarrow} D=5 \stackrel{\phi_+}{\longrightarrow} D=4  \stackrel{y}{\longrightarrow} D=3,
\end{align}
i.e., first a time-like reduction to a Euclidean $D=5$ theory and then two space-like\footnote{For simplicity we use notation $t, \phi_+$, and $y$, to denote directions over which we perform dimensional reduction. It should be kept in mind that only asymptotically this notation is fully justified.} reductions.

The individual steps of the reduction process are fairly standard and we refer to~\cite{Pope} for general reference and to~\cite{Virmani:2012kw,Sahay:2013xda} for calculations very similar to the ones performed here.\footnote{Different orders of reduction of five-dimensional supergravity were recently investigated in~\cite{Cortes:2014jha}.}

\subsection{Time-like reduction to $D=5$}

We make the metric ansatz
\begin{align}
ds_6^2 = -e^{\sqrt{\frac32}\Psi} (dt+ A^{1}_m dx^m)^2 + e^{-\frac{1}{\sqrt{6}} \Psi} ds_5^2
\end{align}
and use standard reduction for the two-form field~\cite{Pope}. The Euclidean five-dimensional theory then contains a Kaluza--Klein vector $A_m^1$ from the metric, one vector field $A_m^2$ from the reduction of the two-form and a five-dimensional two-form. The five-dimensional two-form can be dualised into a vector field according to the relation
\begin{align}
H^{mnp} = \frac12 e^{\sqrt{2}\Phi-\frac{2}{\sqrt{6}}\Psi} \epsilon^{mnpqr} F_{qr}^{3},
\end{align}
where $F^3_{mn}=2\partial_{[m} A_{n]}^3$ is the field strength of the dual vector field. This brings the total number of vector fields up to three that we will label $A_m^I$. The resulting Euclidean $D=5$ theory can be written as
\begin{align}
\label{L5}
\mathcal{L}_{(5)} =  R -\frac12 G_{IJ} \partial_m h^I \partial^m h^J + \frac14 G_{IJ} F^I_{mn} F^{mn\, J} +\frac1{24}C_{IJK} \epsilon^{mnpqr} A^I_m F^J_{np} F^K_{qr},
\end{align}
where we have defined
\begin{align}
h^1 = e^{-\frac{2}{\sqrt{6}}\Psi},\quad
h^2 = e^{\frac1{\sqrt{2}}\Phi+\frac{1}{\sqrt{6}}\Psi},\quad
h^3 = e^{-\frac1{\sqrt{2}}\Phi+\frac{1}{\sqrt{6}}\Psi},
\end{align}
satisfying $h^1h^2h^3=1$ and $G_{IJ}= \delta_{IJ} (h^I)^{-2}$ for $I=1,2,3$. The Chern--Simons terms are defined using $C_{IJK}$ which is totally symmetric, satisfies $C_{123}=1$ and vanishes when two indices are identical. The difference of~\eqref{L5} to the Lorentzian theory that one would have obtained by a space-like reduction (cf.~\cite{Sahay:2013xda}) lies solely in the sign of the kinetic term for the vector fields.

\subsection{Space-like reduction to $D=4$}
\label{app:red4}

The next step is to reduce this theory over a spatial direction to four dimensions. The metric ansatz is\footnote{We use $m$ to label the `non-compact' directions in any reduction step in order not to introduce numerous new index sets. Moreover for simplicity of writing we use $z_5$ instead of $\phi_+$.}
\begin{align}
ds_5^2 = f^2 (dz_5 + A^{0}_m dx^m)^2 + f^{-1} ds_4^2.
\end{align}
The reduction of~\eqref{L5} then is 
\begin{align}
\label{L4}
\mathcal{L}_{(4)} &= R -\frac32 f^{-2} (\partial f)^2 
- \frac12 G_{IJ}\partial_m h^I \partial^m h^J +\frac12 G_{IJ} f^{-2} \partial_m \chi^I \partial^m\chi^J\nn\\
&\quad
+\frac14 G_{IJ} f F_{mn}^I F^{mn\,J}- \frac14 f^{-3} F_{mn}^0 F^{mn\,0} 
+\frac1{8} C_{IJK} \epsilon^{mnpq} \chi^I F_{mn}^J F_{pq}^K\nn\\
&\quad 
+\frac1{8} C_{IJK} \epsilon^{mnpq} \chi^I \chi^JF_{mn}^0 F_{pq}^K
+\frac1{24} C_{IJK} \epsilon^{mnpq} \chi^I \chi^J \chi^K F^0_{mn} F_{pq}^0,
\end{align}
where we have used 
\begin{align}
F_{mn}^{I(4d)} = F^{I(5d)} _{mn} + 2A_{[m}^0 F_{n]z_5}^{I(5d)}
\end{align}
and $F_{n z_5}^{I(5d)} = \partial_n \chi^I$ for some four-dimensional scalar field $\chi^I=A_{z_5}^I$. 

This is a Euclidean $D=4$ theory which falls into the realm of $N=2$ Euclidean supergravity in $D=4$~\cite{Cortes:2003zd,Mohaupt:2007md,Gutowski:2012yb}. The Euclidean $N=2$ formalism is very similar to the standard Lorentzian formalism but uses special para-K\"ahler geometry instead of special K\"ahler geometry. It is based on split complex number $z=x+ e y$ where $x,y\in \mathbb{R}$ and the para-imaginary unit $e$ satisfies $e^2=+1$ and $\bar{e}=-e$. Real and imaginary parts are then defined in the obvious way with respect to $e$. 

The general $N=2$ Euclidean supergravity with vector superfields then has the action
\begin{align}
\label{EuclL4}
{\mathcal L}_{(4)} = R - 2 g_{I \bar{J}} \partial_m X^I \partial^m \bar{X}^{\bar{J}} + \frac{1}{8}\epsilon^{mnpq}  F^\Lambda_{mn}   G_{pq\,\Lambda},
\end{align}
where $\Lambda$ can be either $I$ or $0$. The para-complex scalar fields $X^I$ are contracted using the metric $g_{I \bar{J}} = \partial_I \partial_{\bar{J}} K$ derived from the K\"ahler  potential
\begin{align}
K = - \log \left[ - e (\bar{X}^\Lambda \mathcal{F}_\Lambda - \bar{\mathcal{F}}_\Lambda X^\Lambda)\right],
\end{align}
which is in turn determined by the holomorphic prepotential $\mathcal{F}(X)$ through its derivatives $\mathcal{F}_\Lambda = \partial_{\Lambda} \mathcal{F}$. The prepotential equally determines the matrix
\begin{align}
\mathcal{N}_{\Lambda\Sigma} = \bar{\mathcal{F}}_{\Lambda\Sigma} + 2 e \frac{(\mathrm{Im}\, \mathcal{F} \cdot X)_\Lambda (\mathrm{Im}\, \mathcal{F} \cdot X)_\Sigma}{X \cdot\mathrm{Im}\, \mathcal{F} \cdot X },
\end{align}
where the Hessian is $\mathcal{F}_{\Lambda \Sigma} = \partial_\Lambda \partial_\Sigma \mathcal{F}$. The vector field terms in~\eqref{EuclL4} are then determined by
\begin{align}
G_{mn\, \Lambda} = (\mathrm{Re}\, \mathcal{N})_{\Lambda \Sigma}  F_{mn}^\Sigma + \frac12 (\mathrm{Im}\, \mathcal{N})_{\Lambda \Sigma} \epsilon_{mnpq}   F^{pq\, \Sigma}.
\end{align}
This formalism matches onto our theory~\eqref{L4} by using the prepotential
\begin{align}
\mathcal{F}(X) = - \frac{X^1 X^2 X^3}{X^0},
\end{align}
the gauge $X^0=1$ and the identification 
\begin{align}
\label{Xident}
X^I=x^I-e y^I= -\chi^I-e fh^I.
\end{align}
 For this choice of prepotential $\mathcal{F}$, the real and imaginary parts of $\mathcal{N}$ are
\begin{align}
\mathrm{Re} \, \mathcal{N} &= \left(
\begin{array}{cccc}
 -2 x_1 x_2 x_3 & x_2 x_3 & x_1 x_3 & x_1
   x_2 \\
 x_2 x_3 & 0 & -x_3 & -x_2 \\
 x_1 x_3 & -x_3 & 0 & -x_1 \\
 x_1 x_2 & -x_2 & -x_1 & 0 \\
\end{array}
\right),\nn\\
\mathrm{Im}\, \mathcal{N} &= \left(
\begin{array}{cccc}
 -y_1 y_2 y_3
+\frac{y_2 y_3 x_1^2}{y_1}
+\frac{y_1 y_3 x_2^2}{y_2}
+\frac{y_1 y_2 x_3^2}{y_3} 
& -\frac{x_1 y_2 y_3}{y_1} 
& -\frac{x_2 y_1 y_3}{y_2} 
& -\frac{x_3 y_1 y_2}{y_3} \\
 -\frac{x_1 y_2 y_3}{y_1} 
 & \frac{y_2 y_3}{y_1}
 & 0 & 0 \\
 -\frac{x_2 y_1 y_3}{y_2} & 0 
 & \frac{y_1 y_3}{y_2} & 0 \\
 -\frac{x_3 y_1 y_2}{y_3} & 0 & 0 & 
 \frac{y_1 y_2}{y_3} \\
\end{array}
\right).
\end{align}

\subsection{Reduction to $D=3$ and $\mathrm{SO}(4,4)$ coset model}
\label{subsec:D=3}
The metric reduction ansatz is
\begin{align}
ds_4^2 = e^{2U} (dy + \omega_m dx^m)^2 +e^{-2 U} ds_3^2,
\end{align}
introducing a Kaluza--Klein vector $\omega_m$. The four-dimensional vector fields, written as forms, reduce according to
\begin{align}
A^{\Lambda\,(4d)} = \zeta^\Lambda (dy + \omega_m dx^m) + A^{\Lambda\, (3d)}.
\end{align}
Thus, we obtain a total of five vector fields in $D=3$ that can be dualised to scalar fields according to
\begin{align}
\label{dual1}
- \partial_m \tilde \zeta_\Lambda = \frac12e^{2U}(\mathrm{Im}\,\mathcal{N})_{\Lambda \Sigma} \epsilon_{mnp} (F^{np\,\Sigma} + \zeta^\Sigma F^{np}) + (\mathrm{Re}\,\mathcal{N})_{\Lambda \Sigma} \partial_m\zeta^\Sigma,
\end{align}
for the four vectors $A^\Lambda_m$, with $F^{np}=2\partial^{[n}\omega^{p]}$ the field Kaluza--Klein field strength, and
\begin{align}
\label{dual2}
- \partial_m \sigma = -  e^{4U} \epsilon_{mnp} F^{np} + \tilde \zeta_\Lambda \partial_m\zeta^\Lambda - \zeta^\Lambda \partial_m \tilde\zeta_\Lambda.
\end{align}
for the Kaluza--Klein vector $\omega_m$.

The total Euclidean theory in $D=3$ is then given by
\begin{align}
\mathcal{L}_{(3)} = R  -\frac{1}{2} G_{ab} \partial_m \varphi^a \partial^m\varphi^b.
\end{align}
This is a non-linear $\sigma$-model for sixteen scalar fields of signature $(8,8)$. The metric is given explicitly by
\begin{align}
\label{cmapst} 
 G_{ab}d\varphi^a d\varphi^b &= 4 dU^2 + 4 g_{I \bar{J}}dz^I d\bar{z}^{\bar J} -
 \frac{1}{4}e^{-4U} \left( d\sigma +  \tilde \zeta_\Lambda d \zeta^\Lambda -
 \zeta^\Lambda d \tilde \zeta_\Lambda \right)^2 \\
 &\hspace{-15mm} + e^{-2U}\left[ -(\mathrm{Im} \,\mathcal{N})_{\Lambda \Sigma}d\zeta^\Lambda
 d\zeta^\Sigma + ((\mathrm{Im}\,\mathcal{N})^{-1})^{\Lambda \Sigma} \left( d\tilde
 \zeta_\Lambda +(\mathrm{Re}\,\mathcal{N})_{\Lambda \Xi} d\zeta^\Xi \right)  \left( d\tilde
 \zeta_\Sigma +(\mathrm{Re}\,\mathcal{N})_{\Sigma \Gamma} d\zeta^\Gamma \right) \nonumber \right],
\end{align}
where $z^I=x^I-e y^I$, cf.~\eqref{Xident}. 

As mentioned in the main body of the paper, the $\sigma$-model can be recognised as corresponding to the coset space
\begin{align}
\mathrm{SO}(4,4)/(\mathrm{SO}(2,2) \times \mathrm{SO}(2,2)).
\end{align}
The non-compact form of the denominator group is due to the time-like reduction involved in the compactification process. The groups $\mathrm{SO}(4,4)$ and $\mathrm{SO}(2,2)\times\mathrm{SO}(2,2)$  are defined in~\eqref{GK} in terms of preserved metrics $\eta$ and $\eta'$ given in~\eqref{invmetrics}.
Using the parametrisation of~\cite{Bossard:2009we,KKV2} for the $\mathrm{SO}(4,4)$ generators, the involution fixing the subgroup is defined explicitly by
\begin{subequations}
\begin{align}
&\tilde\tau(H_0) = -H_0, 
& & \tilde\tau(H_I) = - H_I, \\
&\tilde\tau(E_0) = + F_0,
& &\tilde\tau(E_I) = + F_I, \\
&\tilde\tau(E_{q_{0}}) = - F_{q_{0}},
& &\tilde\tau(E_{q_{I}}) = + F_{q_{I}}, \\
&\tilde\tau(E_{p^{0}}) = + F_{p^{0}}, 
& &\tilde\tau(E_{p^{I}}) = - F_{p^{I}}
\end{align}
\end{subequations}
and we define the generalised transpose of an $\mathfrak{so}(4,4)$ Lie algebra element $x$ by
\begin{align}
x^\sharp = -\tilde{\tau}(x)=\eta'x^T \eta',
\end{align}
so that $\mathfrak{so}(2,2)\oplus\mathfrak{so}(2,2)$ elements are $\sharp$-anti-symmetric.

The coset element can be written in Borel gauge as
\begin{align}
\label{cosel}
V = e^{-U H_0} \cdot \left[\prod_{I=1,2,3} \left(e^{-\frac12 \log y^I H_I} e^{-x^I E_I}\right)\right]
\cdot e^{-\zeta^\Lambda E_{q_\Lambda}-\tilde{\zeta}_\Lambda E_{p^\Lambda}} \cdot e^{-\sigma E_0},
\end{align}
and the coset metric then takes the form 
\begin{align}
G_{ab} d\varphi^a d\varphi^b = \mathrm{Tr} (PP)
\quad\textrm{with}\quad
P = \frac12\left( dV V^{-1} + (dV V^{-1})^\sharp\right).
\end{align}

\subsection{Different $\mathrm{SL}(3,\reals)$ vacuum truncations}
\label{shifts}
\textit{Euclidean $D=5$ gravity.} We can perform a truncation to pure $D=5$ Euclidean gravity by imposing
\begin{align}
x^I=0,\quad
y^I=y,\quad
\zeta^I=0,\quad
\tilde{\zeta}_I=0.
\end{align}
The resulting formally $D=6$ metric looks like
\begin{align}
ds_6^2 = -dt^2 + ds_5^2
\end{align}
and $ds_5^2$ is the Euclidean $D=5$ metric. After reduction to $D=3$ over two commuting spatial isometries as above, the metric can be parametrised by the five scalar fields 
\begin{align}
U,\quad
y,\quad
\sigma,\quad
\zeta^0,\quad
\tilde{\zeta}_0.
\end{align}
These form an $\mathrm{SL}(3,\reals)/\mathrm{SO}(1,2)$ subspace of $\mathrm{SO}(4,4)/(\mathrm{SO}(2,2)\times\mathrm{SO}(2,2))$. The $\mathfrak{sl}(3,\reals)$ Lie algebra is generated by 
\begin{align}
H_0, \quad
H_1+H_2+H_3,\quad
E_{q_0},\quad
E_{p^0},\quad
E_0
\end{align}
and their transposes.

\textit{Lorentzian $D=5$ gravity.} Alternatively, we can embed a Lorentzian five-dimensional metric in the six-dimensional theory according to
\begin{align}
ds_6^2 = dy^2 + d\tilde{s}_5^2,
\end{align}
where $d\tilde{s}_5^2$ is now a Lorentzian $D=5$ metric. Note that here we have singled out the $y$-coordinate that---in the general parametrisation above---was associated with the reduction from $D=4$ to $D=3$. Analysing metrics of this form we find that they can be parametrised by the following five independent scalar fields:
\begin{align}
\tilde{\zeta}_0,\quad
\tilde{\zeta}_1,\quad
\chi^1,\quad
y^1= f^3 e^{-4U},\quad
y^2=y^3=e^{2U}.
\end{align}
In terms of $\mathrm{SO}(4,4)$ generators the corresponding $\mathrm{SL}(3,\reals)$ subgroup is now generated by
\begin{align}
H_1, \quad
H_0+H_2+H_3,\quad
F_{p^1},\quad
E_{p^0},\quad
E_1
\end{align}
and their transposes. Note that the intersection of this $\mathrm{SL}(3,\reals)$ with $\mathrm{SO}(2,2)\times\mathrm{SO}(2,2)$ also yields an $\mathrm{SO}(1,2)$ subgroup.

\textit{Relation between the two truncations.} The two $\mathrm{SL}(3,\reals)$ subgroups discussed for the Euclidean and Lorentzian truncation are related by a conjugation in $\mathrm{SO}(4,4)$. This conjugation is by the element
\begin{align}
w = e^{\frac{i\pi}{2} (E_{q_2}+E_{q_2}^\sharp) +\frac{i\pi}{2} (E_{q_3}+E_{q_3}^\sharp)}
\end{align}
and explains the shift by $\frac{i\pi}{2}$ in the charging parameter when we work with the MP instanton rather than an overrotating black hole.

\section{Detailed intermediate expressions}
\label{app:details}

In this appendix we present the details of some of the intermediate expressions that enter the derivation of the JMaRT fuzzball.

\subsection{Values of $M(x)$ from inverse scattering}
\label{app:Mvalues}

Factorising the monodromy $\cM(w)$ of~\eqref{M11} according to the inverse scattering procedure leads to the following non-trivial components of the spacetime matrix $M(x)=m_{ab}$,
\begin{subequations}
\begin{align}
m_{33} &= -\frac{2 \left(2 \zeta_{11} (u+1)-\zeta_{22}^2
   (u+v)\right)}{\left(\zeta_{22}^2-2 \zeta_{11}\right) \left(2
   \zeta_{11} \left(1-u^2\right)+\zeta_{22}^2
   \left(u^2-v^2\right)\right)}, \\ 
m_{44} &= \frac{\zeta_{11}^2 \left(4 u^2+8 u+4\right)+2 \zeta_{11}
   \zeta_{22}^2 \left(-2 u^2-2 u+v^2+2 v+1\right)+\zeta_{22}^4
   \left(u^2-v^2\right)}{\left(\zeta_{22}^2-2 \zeta_{11}\right)
   \left(2 \zeta_{11} \left(1-u^2\right)+\zeta_{22}^2
   \left(u^2-v^2\right)\right)},  \\
m_{35} &= \frac{-2 \zeta_{11} \zeta_{22}^2+u^2 \left(
\zeta_{22}^2-2 \zeta_{11}\right)^2-2 u \left(\zeta_{11}
   \zeta_{22}^2-2 \zeta_{11}^2\right)-\zeta_{22}^2 v^2
   \left(\zeta_{22}^2-2 \zeta_{11}\right)+2 \zeta_{11}
   \zeta_{22}^2 v}{\left(\zeta_{22}^2-2 \zeta_{11}\right) \left(2 \zeta_{11} \left(1-u^2\right)+\zeta_{22}^2 \left(u^2-v^2\right)\right)},  \\
m_{47}&=-\frac{2 \left(2 \zeta_{11}^2 \zeta_{22}+u \left(4 \zeta_{11}^2 \zeta_{22}-2 \zeta_{11} \zeta_{22}^3\right)+v
   \left(2 \zeta_{11} \zeta_{22}^3-2 \zeta_{11}^2
   \zeta_{22}\right)\right)}{\left(\zeta_{22}^2-2 \zeta_{11}\right) \left(2 \zeta_{11} \left(1-u^2\right)+\zeta_{22}^2 \left(u^2-v^2\right)\right)},  \\
m_{41} &= \frac{2 (-2 \zeta_{11} \zeta_{22}-2 \zeta_{11}
   \zeta_{22} v)}{\left(\zeta_{22}^2-2 \zeta_{11}\right)
   \left(2 \zeta_{11} \left(1-u^2\right)+\zeta_{22}^2
   \left(u^2-v^2\right)\right)}.
\end{align}
\end{subequations}
{}From this one can reconstruct the scalar fields of the Myers--Perry instanton using the formulas~\eqref{scalarsFromM} and~\eqref{eq:sig}.

\subsection{Three-dimensional one-forms after dualisation}
\label{app:dualone}

Dualising the relevant scalar fields of the $\mathrm{SO}(4,4)$ coset element after the charging transformations leads to the following one-forms in $D=3$,
\bea
\omega_3 &=& \frac{2 c \zeta_{22} s_2 s_3 \left(u^2
   \left(\zeta_{22}^2-2 \zeta_{11}\right)+2 \zeta_{11} u
   \left(v^2-1\right)+v^2 \left(2 \zeta_{11}-
\zeta_{22}^2\right)\right)}{\left(2 \zeta_{11}-\zeta_{22}^2\right)
   \left(2 \zeta_{11} \left(u^2-1\right)+\zeta_{22}^2
   \left(v^2-u^2\right)\right)} dz_3, \\
A_3^0&=&
\frac{2 c \left(-\left(u^2-1\right) v \left(2 
\zeta_{11}-\zeta_{22}^2\right)-\zeta_{22}^2 u+\zeta_{22}^2
   u v^2\right)}{\left(2 \zeta_{11}-\zeta_{22}^2\right) \left(2
   \zeta_{11} \left(u^2-1\right)+\zeta_{22}^2
   \left(v^2-u^2\right)\right)} dz_3, \\
A_3^1&=&-\frac{2 c \zeta_{22} c_2 c_3 \left(u^2 \left(\zeta_{22}^2-2
   \zeta_{11}\right)-2 \zeta_{11} u \left(v^2-1\right)+v^2
   \left(2 \zeta_{11}-
\zeta_{22}^2\right)\right)}{\left(\zeta_{22}^2-2 \zeta_{11}\right)
   \left(\zeta_{22}^2 \left(u^2-v^2\right)-2 \zeta_{11}
   \left(u^2-1\right)\right)} dz_3,\\
A_3^2&=&\frac{2 c \zeta_{22} c_2 s_3
   \left(-2 \zeta_{11} u^2+\zeta_{22}^2 u^2-2 \zeta_{11}
   u+2 \zeta_{11} u v^2+2 \zeta_{11} v^2-\zeta_{22}^2
   v^2\right)}{\left(\zeta_{22}^2-2 \zeta_{11}\right) \left(2
   \zeta_{11}-2 \zeta_{11} u^2+\zeta_{22}^2
   u^2-\zeta_{22}^2 v^2\right)} dz_3,\\
A_3^3&=&\frac{2 c \zeta_{22} s_2 c_3
   \left(-2 \zeta_{11} u^2+\zeta_{22}^2 u^2-2 \zeta_{11}
   u+2 \zeta_{11} u v^2+2 \zeta_{11} v^2-\zeta_{22}^2
   v^2\right)}{\left(\zeta_{22}^2-2 \zeta_{11}\right) \left(2
   \zeta_{11}-2 \zeta_{11} u^2+\zeta_{22}^2
   u^2-\zeta_{22}^2 v^2\right)} dz_3.
\eea

\end{document}